\documentclass[a4paper, 11pt,english]{article}

\usepackage[T1]{fontenc}

\usepackage{amsmath}
\usepackage{amstext}
\usepackage{wasysym}
\usepackage{esint}
\usepackage{babel}
\usepackage{lscape}
\usepackage{cancel}
\usepackage{bm}
\usepackage{bbm}
\usepackage{multirow} 
\usepackage{amsfonts}
\usepackage{amssymb}
\usepackage{graphicx}
\usepackage[usenames,dvipsnames]{color}
\usepackage{hyperref}
\usepackage{epstopdf,epsfig}
\usepackage{verbatim}
\usepackage[utf8]{inputenc}
\usepackage[left=1in,right=1in,top=1in,bottom=1in]{geometry}             % For good copy.
\usepackage{booktabs} % for much better looking tables
\usepackage{array} % for better arrays (eg matrices) in maths
\usepackage{paralist} % very flexible & customisable lists (eg. enumerate/itemize, etc.)
\usepackage{verbatim} % adds environment for commenting out blocks of text & for better verbatim
\usepackage{subfig} % make it possible to include more than one captioned figure/table in a single float
\hypersetup{
    colorlinks=true,         % false: boxed links; true: colored links, false is default
    linkcolor=blue,          % color of internal links, red is default
    citecolor=red,        % color of links to bibliography, 'green' is default
    urlcolor=Violet             % color of external links, cyan is default
}

\begin{document}

\title{General Covariance from the Quantum Renormalization Group}
\author{{\bf Vasudev Shyam}\footnote{\href{mailto:vshyam@pitp.ca}{vshyam@pitp.ca}} \\\it Perimeter Institute for Theoretical Physics\\ \it 31 N. Caroline St. Waterloo, ON, N2L 2Y5, Canada }
\date{\today}
\maketitle

\begin{abstract}
The Quantum renormalization group (QRG) is a realisation of holography through a coarse graining prescription that maps the beta functions of a quantum field theory thought to live on the `boundary' of some space to holographic actions in the `bulk' of this space. A consistency condition will be proposed that translates into general covariance of the gravitational theory in the $D + 1$ dimensional bulk.  This emerges from the application of the QRG on a planar matrix field theory living on the $D$ dimensional boundary. This will be a particular form of the Wess--Zumino consistency condition that the generating functional of the boundary theory needs to satisfy. In the bulk, this condition forces the Poisson bracket algebra of the scalar and vector constraints of the dual gravitational theory to close in a very specific manner, namely, the manner in which the corresponding constraints of general relativity do. A number of features of the gravitational theory will be fixed as a consequence of this form of the Poisson bracket algebra. In particular, it will require the metric beta function to be of gradient form.  

\end{abstract}

\section{Motivations and Introduction}

 The Quantum Renormalization group (QRG) due to Sung-Sik Lee \cite{S.S.Lee1}, \cite{S.S.Lee2}, is a constructive coarse graining prescription used for finding holographic dual descriptions of quantum field theories. In this article, I will propose a consistency condition that this coarse graining procedure needs to satisfy in order for the holographically dual theory to possess general covariance. 
 
In order to focus on this feature of covariance in the emergent dual theory, it will help to make a specific choice of the field theory on which the application of the QRG results in the emergence of a semiclassical theory involving only a dynamical metric tensor (or in other words, pure gravity). The rules of effective field theory dictate that the problem of picking out general relativity from the space of all possible theories of a dynamical metric tensor reduces entirely to ensuring diffeomorphism invariance and having two derivatives in the action. From this perspective, these consistency conditions which are `dual' to general covariance, are a key prerequisite that the hypothetical holographic dual to general relativity, should such a theory exist, need satisfy. 

 First, some explanation is necessary about what the QRG procedure entails and what theory to apply it to so that pure gravity emerges in the bulk. The basic idea is to map the renormalization group flow of a quantum field theory into the dynamics of a dual theory living in one dimension higher, where the additional direction is equated with the energy scale of the quantum field theory. The RG flow equations describe how coupling constants of operators in a given quantum field theory run under change of scale. If these are to be equated with dynamical field equations, and fields can vary as functions of space, then the couplings of operators should be promoted to space dependent sources. This idea goes under the name of the local renormalisation group (LRG), and can be seen as the QRG's precursor. Here couplings all of operators (composite operators included) in the action of the theory under consideration are upgraded to space dependent, background sources. The coarse graining transformations are implemented through local Weyl transformations of the background metric, which too is now arbitrary and space dependent. The LRG equations describe the response of the effective action and correlation functions to such transformations. 

Then, the QRG is a prescription that assigns dynamics to a suitable subset of these sources. The `evolution parameter' of these dynamical sources is the RG scale (or RG time). The dependence of these sources on the RG time in addition to the already present space dependence implies that this dual theory lives in one higher dimension than the field theory whose RG flow it encodes. This is how the QRG realises holography, and how the emergent direction of space is equated with the RG time. The limit in which the dual theory becomes semi-classical will be of primary interest in what is to follow. 
 
Most quantum field theories of interest possess an energy momentum tensor, which is a composite operator whose source is the background metric. The local renormalization group idea of making all sources arbitrary and space dependent, metric included, therefore requires putting the theory on an arbitrary background geometry. If the energy momentum tensor is contained in the subset of operators whose sources the QRG will assign dynamics to, then the dual theory will involve a dynamical metric tensor.

Then to obtain pure gravity in the bulk, the Quantum renormalization group has to somehow focus only on the energy momentum tensor, so one assumes a limit where no other operators are generated through quantum corrections. In other words, it must be assumed that only the energy momentum tensor of the theory possesses finite scaling dimension. The existence of such a limit is not guaranteed, and is indeed an assumption. In order to be more concrete, the field theory to which QRG and this limit are applied must be specified. 

The claim in \cite{S.S.Lee1} is that  the application of QRG and the aforementioned limit to a {\it matrix field theory} in the 'tHooft limit results in the emergence of a semiclassical theory of pure gravity. The planarity of the matrix field theory corresponds to the taking the semiclassical limit of the dual theory, and so, the bulk theory is given in the low energy limit. The infinite gap in the spectrum of scaling dimensions, {\em i.e.} the gap between the scaling dimension of the energy momentum tensor and that of all other operators, is responsible for the metric being the only source rendered dynamical through the QRG. This specifies the field content of the bulk theory. Then, covariance and possessing two derivatives are the only remaining conditions that need to be satisfied in order to force this low energy bulk theory to be general relativity. 

In \cite{S.S.Lee1}, \cite{S.S.Lee2}, it was shown that the bulk theory which emerges from the application of QRG to the aforementioned theory has two derivatives in the emergent radial direction. However, it possessed an arbitrary number of spatial derivatives, i.e. derivatives along the directions of the space in which the planar matrix field theory lives. It was assumed then, that there exists a long wavelength limit where the number of spatial gradients when arranged in a gradient expansion could somehow be truncated in an appropriate manner so that the theory is covariant to leading order in this expansion. This theory would then be general relativity. Any of the higher order terms in the gradient expansion would necessarily break general covariance. 

The problem here is that demanding suppression of these non covariant corrections beyond leading order in said gradient expansion requires appealing to some notion of strong coupling (which is described in \cite{NakayamaLRG} for instance) in addition to the demand that the hypothetical dual theory possess a large gap in the spectrum of anomalous dimensions. This is poorly understood, and perhaps too much to demand even from a hypothetical theory. In this article, I will make the case that the situation can be significantly improved upon by considering the consistency conditions the local renormalization group need satisfy. These are known as the Wess--Zumino consistency conditions and I will propose a particular manner in which they need to be satisfied so that the bulk theory obtained from the QRG procedure is manifestly covariant. 
The aforementioned condition I'm proposing, which can be considered the `holographic' Wess--Zumino consistency condition takes the form:
\begin{equation}[\Delta_{\sigma},\Delta_{\sigma'}]W[g]=\int \textrm{d}^{3}x \sqrt{g} g^{\mu\nu}\left(\sigma\partial_{\mu}\sigma'-\sigma'\partial_{\mu}\sigma\right)\langle \nabla_{\rho}T^{\rho}_{\nu} \rangle.\end{equation}
The left hand side represents the commutator of local RG transformations (denoted $\Delta_{\sigma}$) acting on the generating functional $W[g]$ of the dual theory, and $T_{\mu\nu}$ on the right hand side is the energy momentum tensor of the boundary field theory. 
This is, in other words an additional consistency condition that the coarse graining procedure itself must satisfy. The nature of this condition is such that in order to satisfy it, the gradient expansion must be truncated in a manner that respects general covariance. 

\ This consistency condition will also fix the functional dependence of the beta functions of the boundary matrix field theory on the metric tensor. In particular, from it will follow the gradient flow property of the metric beta function, {\em i.e.} 
\begin{equation}
\beta_{\mu\nu}(g)=G_{\mu\nu\alpha\beta}\frac{\delta c[g]}{\delta g_{\alpha\beta}},
\end{equation}
where the functional $c[g]$ can be seen as the analogue of the running central charge of the boundary field theory. The tensor $G_{\mu\nu\alpha\beta}$, known as the de-Witt super metric, is analogous to the Osborn-Zamalodchikov metric on theory space. This form of the beta function following from the Wess--Zumino consistency conditions is exactly analogous to what happens in studies of the local renormalization group in the literature: \cite{Osborn1}, \cite{Freidan1}. 

\section{Renormalization Group flow as dynamics on the phase space of sources}

 This semiclassical limit of the dual theory obtained from the mapping the QRG provides is a precise sense in which the renormalization group flow of the boundary quantum field theory can be seen as dynamical evolution of the sources in RG time. The renormalization group flow equations most often encountered are those that define beta functions for the quantum field theory under consideration and these equations are first order differential equations. If they are to be recast as the equations of motion for some dynamical theory of the sources, then the natural expectation is that these RG flow equations are seen as Hamilton's equations of the dual theory, provided an appropriate identification of the phase space and the conjugate variables on it is made. In other words the RG flow is identified with the flow generated by some Hamiltonian on the appropriate phase space of the theory of dynamical sources. 
 
 \subsection{The case of the local renormalization group}
 
Before delving into what variables ought to parameterise the phase space mentioned in the introduction, it will help to first review some rudimentary aspects of the local renormalization group. For a more detailed pedagogical introduction, see for instance chapter 3 of \cite{NakayamaScalevsConformal} . The central object of interest will be the renormalised generating functional given by the logarithm of the partition function in the presence of external sources (including those which couple to composite operators). This object is denoted $W[J]$ and is given by \footnote{ Indices such as $I$ are internal indices. For simplicity, in this section I will deal with solely with spacetime scalar operators.}
\begin{equation}e^{-W[J^{I}(x)]}=\int \mathcal{D}\Phi\, e^{-S[\Phi;J^{I}(x)]},\end{equation}
were the sources $\left\{J^{I}(x)\right\}$ couple to operators $\left\{\mathcal{O}_{I}(x)\right\}.$ The definition of correlation functions of local composite operators is given by taking functional derivatives of the generating functional with respect to the appropriate sources, for example, the one point functions of local operators can be computed via:
\begin{equation}\langle\mathcal{O}_{I}(x)\rangle=\frac{\delta W}{\delta J^{I}(x)}.\end{equation}

I will assume that there aren't any intrinsic mass scales in the theory. This means that scales are generated by quantum effects alone. The coarse graining method employed in local RG involves performing local Weyl transformations of the background metric
$$g_{\mu\nu}\rightarrow e^{2\sigma(x)}g_{\mu\nu}.$$
This is a geometrical generalisation of Kadanoff's blocking transformations which are applicable in real space renormalization group applied to lattice models to continuum quantum field theories. The local renormalization group is the study of the response of the generating functional to such Weyl transformations. 

More precisely, the local Callan--Symanzik equation is given by
\begin{equation}\Delta_{\sigma}W[J^{I}(x)]\equiv \int \textrm{d}^{D}x \sqrt{g}\sigma(x)\left(g_{\mu\nu}\frac{\delta}{\delta g_{\mu\nu}}-\beta^{I}_{J}(x)\frac{\delta }{\delta J^{I}(x)}\right)W=\mathcal{A}_{\sigma}[J].\end{equation}
This equation is valid in a neighbourhood of the fixed point of the renormalization group. 
The  $\beta^{I}_{J}(x)$ are beta functions which encode the running of the space dependent sources. Note that this collection of sources includes the metric, so there will be a term of the form $\beta_{\mu\nu}\frac{\delta}{\delta g_{\mu\nu}}$ such that the first $g_{\mu\nu}\frac{\delta}{\delta g_{\mu\nu}}$ term can be seen as the lowest order in derivative part. The reason for it being stripped away from the other terms in the metric beta function is that the above form of the local Callan--Symanzik equations make the geometrical role of $\Delta_{\sigma}$ generating Weyl transformations clearer. The term $\mathcal{A}_{\sigma}$ is the integrated conformal anomaly, smeared against Weyl factor $\sigma(x)$. The n-point functions of the field theory under consideration can be computed through functional differentiation of the generating functional with respect to sources. Then, the above local Callan--Symanzik equation will dictate how these n-point functions ought to run under a change of scale. The more traditional form of these equations can be recovered by taking the limit where the background geometry is flat.
Now, the objective is to gain intuition for how the above equations can be recast into Hamilton's equations, and thereby identify what the phase space for such Hamiltonian evolution ought to be. 
Consider the limit of the above expression where the metric is taken to be flat, and the local Weyl transformations are replaced by global dilatations. The above equation in this limit takes the form
\begin{equation}\frac{\partial W}{\partial \textrm{ln}\mu} =-\beta^{I}\langle\mathcal{O}_{I}\rangle. \end{equation}
The anomaly term depends on the derivatives of the metric tensor at least to second order (except for a cosmological constant term which would play a role should we include massive couplings in the theory) and hence it vanishes in the aforementioned limit. 
 Dolan in \cite{Dolan} proceeded to posit that the sources $\left\{J^{I}(x)\right\}$ can be seen as though they were canonically conjugate to the vacuum expectation values of the renormalised operators to which they couple $\left\{\langle \mathcal{O}_{I}(x)\rangle\right\}\equiv\left\{p_{I}(x)\right\}.$ The evolution parameter is the RG time $\mu$ which is the global remnant of $\sigma(x)$. The generating functional can be seen as Hamilton's principal function in which case, the Hamiltonian can be identified directly form the global generalisation of the local Callan--Symanzik equations as: 
\begin{equation}H_{RG}=p_{I}\beta_{J}^{I}.\end{equation}
Thus solving every one of Hamilton's equations will be equivalent to solving all of the renormalization group flow equations. This will not be a practical to do in the case of strongly coupled field theories where one would have to solve an infinite number of such equations. Also notice that this Hamiltonian is linear in the momentum, so the evolution it generates is rather simple as it is that of a system whose kinetic energy is entirely quenched\footnote{There are sufficiently non-trivial contexts such as higher spin-holography where the RG Hamiltonian takes a form similar to this, see \cite{LeighRG} }. If a term that is quadratic at the very least in the momentum could somehow be generated, then the situation would be more similar to Hamiltonian evolution most often encountered in mechanics or even in field theories. The question then would be what such evolution describes on the other side of the duality. This is what the quantum renormalization group addresses. 

\subsection{The semi-classical limit of the bulk theory in QRG}

This subsection will present the Quantum Renormalization group flow of the planar matrix field theory. This example will be of interest because the QRG Hamiltonian defined on the phase space of sources and their conjugate vacuum expectation values, whose structure is dictated by the form of quantum corrections to the seed action in one step of RG, will be quadratic in the momenta. This was first worked out in \cite{S.S.Lee2} and similar results were derived in \cite{Akhmedov} and in \cite{Imbimbo2} under the name of the `planar Polchinski equation'. The fundamental fields are Hermitian $N\times N$ matrix
fields $\Phi(x)$ (I will suppress the matrix indices for notational ease). Assume also that the matrix model is gauge
under some gauge group ($U(N)$ for instance), this means that gauge
invariant operators in the action are necessarily sums and products
of traces of monomials of the matrix fields and their derivatives.

The simplest of such operators are the so called single trace operators, given by:

\begin{equation}
O_{\left\{ m\right\} }=\frac{1}{N}\sqrt{g}\textrm{tr}\left(\Phi(\nabla_{\mu_{1}^{1}}\cdots\nabla_{\mu_{p_{1}}^{1}}\Phi)\cdots(\nabla_{\mu_{1}^{q}}\cdots\nabla_{\mu_{p_{q}}^{q}}\Phi)\right),
\end{equation}
where the multi index set $\{\mu_{p_{i}}^{i}\}$ is used to denote
the fact that there can be varying number of derivatives and arbitrary
permutations of indices thereof in each term of the product of derivatives
in the above operator. For brevity, I will use just a single latin
letter index $m$ to encapsulate the multi-index set above. Multi-trace
operators are formed from products and derivatives of products of
single trace operators. The sources for single trace
operators will be denoted as $J^{m}$ and for multi trace operators as $\mathcal{J}^{m}$.
These sources have arbitrarily many indices and in particular,
the source for the term in the action with two derivatives can be
chosen to be the background metric.

 The action reads:

\begin{equation}
S=S_{o}[\Phi(x)]+N^{2}\sum_{m}\int\textrm{d}^{3}x\sqrt{g}\underset{Single-Trace}{\underbrace{J^{m}(x)O_{m}}}+\sum_{m}\int\textrm{d}^{3}x\underset{Multi-Trace}{\sqrt{g}\underbrace{V_{m}[O_{m},\mathcal{J}^{m}]}}.
\end{equation}
 For reasons which will be made clear in the remainder of this sub-section,
it will be of interest to study the renormalization group trajectories starting
from the sub space of single trace operators. The generating functional
for a theory on this sub space is given by

\begin{equation}
Z[J^{m}]=\varint D\Phi\exp i\{S[O_{m}(\Phi);J^{m}]\}.
\end{equation}

Wilsonian RG describes how the sources(/couplings) change when the
UV cutoff $\Lambda$ is lowered by a factor\footnote{This infinitesimal `RG time step' $\delta z$ is introduced to keep track of how many iterations of infinitesimal local RG transformations one performs.} $\sigma(x)\equiv \alpha(x)\delta z$:
$$\Lambda\rightarrow e^{\alpha(x)\delta z}\Lambda,$$ 
through the equations defining beta functions $\beta^{m}$ for sources $J^{m}(x)$.
Solutions to those first order ODE's will determine a path in the space
of sources/couplings. As mentioned before, when the theory is strongly coupled, there are,
in general, infinitely many such sources. In that case, solving all RG flow equations
becomes intractable.

Now, in the limit as $N\rightarrow\infty$, the quantum corrections to the single trace action
under one step of RG ($\Lambda\rightarrow \Lambda-\Lambda\delta z\alpha(x))$ take the form\footnote{the indices in braces such as $\left\{\mu\right\}, \left\{\nu\right\}$ etc. of differential operators and tensors denote multi index sets distinct from those incorporated into the indices $m,n$ and will be used solely in situations where a differential operator of arbitrarily high order is involved. For instance, $A^{m\left\{\mu\right\}}\nabla_{\left\{\mu\right\}}\equiv \sum^{k_{m}}_{k=0}A^{m \mu_{1}\cdots \mu_{k}}\nabla_{\mu_{1}}\cdots\nabla_{\mu_{k}}$, and $k_{m}$ could in principle be infinity. } :

\[
\delta S[O_{n},J^{n}]=N^{2}\int\textrm{d}^{D}x\sqrt{g}\alpha(x)\delta z(\mathcal{L}_{C}(J^{n}(x))-\beta^{m}(J^{n}(x))O_{m}+
\]

\begin{equation}
\frac{G^{mn\left\{ \mu\right\} \left\{\nu\right\}}(J^{n}(x))}{2}\nabla_{\left\{ \mu\right\} }O_{m}\nabla_{\left\{ \nu\right\} }O_{n})+\mathcal{O}(\delta z^{2})
\end{equation}

Here, the term $\mathcal{L}_{C}(J^{n}(x))$ is the integrand within the anomaly term denoted in the previous subsection as $\mathcal{A}_{\sigma}$, {\em i.e.} $\int \textrm{d}^{D}x\sqrt{g} \delta z\alpha(x)(\mathcal{L}_{C}(J^{n}(x))\equiv\mathcal{A}_{\sigma=\alpha(x)\delta z}$.

The value of applying quantum RG to the matrix field
theory will become apparent here, because it cleverly re-organises the renormalization
of this theory without having to ever leave the subspace of single
trace operators. The honest fixed point of the flow cannot be projected down to
this subspace due to the fact that multi trace operators are generated by quantum corrections. QRG is a method to nevertheless restrict the RG trajectory to this
subspace and thereby project down the fixed point, by paying the price
of promoting the sources in this subspace to fluctuating quantum fields.
This is why it is apt to call this technique the quantum renormalization
group. I will flesh out what is meant by the above statement schematically in what follows.

The most important thing to notice about the quantum corrections
to the single trace action is that to linear order in $\delta z$, only
double trace operators are generated.
\subsection*{Toy integral example}
Making precise the statement about the single trace sources being promoted to dynamical fields lies in the observation that the multi-trace operators (double trace in the planar limit) can be removed by paying the price of functional integration over auxiliary fields. To schematically describe this, I will take the example of an ordinary (as opposed to functional) integral over a single variable for simplicity. The main idea will carry through into the case of interest. Let the toy integrand to be transformed be of the following exponential form:
\begin{equation}\int\, \textrm{d}\varphi e^{iJO}=Z(J)=\int \textrm{d}\varphi\, e^{i\left\{JO(\varphi)+\alpha \delta z(\beta(J) O(\varphi)+G(J)O^{2}(\varphi)+l_{C})\right\}}\end{equation}
Notice that each of these terms is analogous to the functional integrand of the single trace planar matrix field theory. The variable $\varphi$ now plays the role of the fundamental field and the source for single trace operators $O(\varphi)$ is the variable $J$. The first term in the exponential is analogous to the single trace action, the second term is analogous to the single trace beta function term and the third to the double trace term. The latter two terms are generated by quantum corrections in one step of RG, in addition to the renormalization of the coupling of the identity or the `cosmological constant' term $l_{C}$, and hence a factor of $\alpha \delta z$ is retained as a reminder of this fact. 

The goal is to remove the quadratic term in the exponent in the integrand at the cost of introducing integration over a new `field' $p^{(1)}$:
\begin{equation}e^{i\left\{JO(\varphi)+\alpha \delta z(\beta(J) O(\varphi)+G(J)O^{2}(\varphi)+l_{C})\right\}}=\int \textrm{d}p^{(1)}\delta(p^{(1)}-O)e^{i\left\{-Jp^{(1)}-\alpha \delta z(\beta(J)p^{(1)}-p^{(1)}G(J)p^{(1)}+l_{C})\right\}}.\end{equation}
The delta function itself can be represented in integral form over another variable $j^{(1)}$:
$$\int \textrm{d}p^{(1)}\delta(p^{(1)}-O)e^{i\left\{-Jp^{(1)}-\alpha \delta z(\beta(J)p^{(1)}-p^{(1)}G(J)p^{(1)}+l_{C})\right\}}$$ $$=\int \textrm{d}p^{(1)}\textrm{d}j^{(1)}e^{i(p^{(1)}-O)j^{(1)}}e^{i\left\{-Jp^{(1)}-\alpha \delta z(\beta(J)p^{(1)}-p^{(1)}G(J)p^{(1)}+l_{C})\right\}}$$
\begin{equation}=\int \textrm{d}p^{(1)}\textrm{d}j^{(1)}e^{ip^{(1)}(j^{(1)}-J)}e^{i(j^{(1)}O)}e^{i\left\{-\alpha \delta z(\beta(J)p^{(1)}-p^{(1)}G(J)p^{(1)}+l_{C})\right\}},\end{equation}
and if the $p^{(1)}$ integral were now performed, one would see that the above expression is hiding a delta function $\delta(j^{(1)}-J)$, which becomes apparent if the above expression is written as:
$$\int \textrm{d}p^{(1)}\textrm{d}j^{(1)}e^{ip^{(1)}(j^{(1)}-J)}e^{i(j^{(1)}O)}e^{i\left\{-\alpha \delta z(\beta(J)p^{(1)}-p^{(1)}G(J)p^{(1)}+l_{C})\right\}}.$$
The fields $j^{(1)}$ and $p^{(1)}$ are denoted in a manner which suggests tentatively that they shall be related to the fluctuating sources and their conjugate vacuum expectation values on the RG phase space. The nature of this relation will be made explicit in what follows. 

So now the `partition function' can be written as
\begin{equation}Z(J)=\int \textrm{d}p^{(1)}\textrm{d}j^{(1)}e^{ip^{(1)}(j^{(1)}-J)}e^{i\left\{-\alpha \delta z(\beta(J)p^{(1)}-p^{(1)}G(J)p^{(1)}+l_{C})\right\}}Z(j^{(1)}),\end{equation}
where $Z(j^{(1)})=\int\textrm{d}\varphi \exp{i\left\{jO\right\}},$ and is thus the same as $Z(J)$ except that $J$ is replaced by $j^{(1)}$. The RG transformations still involve coarse graining with respect to the fundamental fields $\varphi$ and so since the quantum corrections have all been factored out into the integral in front of $Z(j^{(1)})$, the second step of RG will proceed exactly the way the first step did except that $J$ is now replaced by the source $j^{(1)}$. Given this guarantee of maintenance of the form of the quantum corrections, one once again may choose to integrate in auxiliary fields $(p^{(2)},j^{(2)})$ to obtain a result similar to the one above, {\em i.e.}
\begin{equation}Z(j^{(1)})=\int \textrm{d}p^{(2)}\textrm{d}j^{(2)}e^{ip^{(2)}(j^{(2)}-j^{(1)})}e^{i\left\{-\alpha^{(2)} \delta z(\beta(j^{(1)})p^{(2)}-p^{(2)}G(j^{(1)})p^{(2)}+l_{C})\right\}}Z(j^{(2)}).\end{equation}
Here, $\alpha^{(2)}$ is the analogue of the Weyl factor chose at the second step of RG, which is free to be chosen to be different from $\alpha$. 
Thus a pattern emerges, and if so the result of iterating this procedure for $k$ RG steps can be written as:
$$ Z(J)=\int \prod^{k}_{i=0}[\textrm{d}\alpha^{(i)}\textrm{d}p^{(i)}\textrm{d}j^{(i)}]\times $$ \begin{equation}\left(e^{i\sum^{k}_{i=0}k\left\{\delta z\left(\frac{p^{(i)}(j^{(i)}-j^{(i-1)})}{\delta z}\right)-\alpha^{(i)}\left(\beta(j^{(i-1)}) p^{(i)}+p^{(i)}G(j^{(i-1)})p^{(i)}+l_{C}\right)\right\}}\vert_{(j^{(0)},p^{(0)})=(J,O)}\right)Z(j^{(k)}).\end{equation}
Note that the different Weyl factors at each step of RG denoted $\alpha^{(i)}$ are also integrated over in order to `average' over all possible RG paths. The consequences of the path independence of this RG procedure will play a very important role in the remainder of this article.

The continuum limit of the above product of integrals can the be defined by sending $\delta z\rightarrow 0$, and defining $z=\epsilon\exp{(k\delta z)}$ as the so called `radial' time. Here, $\epsilon$ denotes a short distance cutoff. The integration variables then become: $\left\{\alpha^{(i)},j^{(i)},p^{(i)}\right\}\rightarrow\left\{\alpha(z),j(z),p(z))\right\}$. The latter two parameterise the dynamical phase space alluded to in earlier discussions.
Then the set of integrals over all RG steps can be recast as the functional integral:
\begin{equation}Z(J)=\int \mathcal{D}\alpha(z)\mathcal{D}j(z)\mathcal{D}p(z)e^{i\int^{z=z_{*}}_{z=0} \textrm{d}z \left(p(z)\frac{\textrm{d}j(z)}{\textrm{d}z}-\alpha(z) H_{QRG}\right)}\vert_{(j(0),p(0))=(J,O)}Z(j(z_{*}))\end{equation}
These `fields' now gaining additional dependence on the RG time is the signature of the emergence of a new directions of space, or in other words, of holography. The field theory of these sources is holographically dual to the original theory of the $\varphi$ fields.  
The bound on the integral $z_{*}$ denotes where the RG transformations are truncated, which needn't necessarily be infinity.  The function $H_{QRG}$ is then give by
\begin{equation}H_{QRG}=\beta(j(z))p(z)+p(z)G(j(z))p(z)+l_{C}.\end{equation}
As promised, the Hamiltonian is now quadratic in the `momenta' $p(z)$ that are conjugate to the dynamical source variables $j(z)$. 
\subsection*{Back to the Matrix Field Theory}
Similarly, in the planar matrix field theory case, the Quantum Renormalization promotes the sources $J^{m}(x)$ and the vacuum expectation values of the single trace operators $\langle O_{m}(\Phi(x))\rangle $ to the dynamical fields $(j^{m}(x,z),p_{m}(x,z))$. In the large $N$ limit, these single trace operators are equal to their vacuum expectation values. These fact that they are labeled by the RG time $z$ in addition to the labels $x$ is the precise sense in which they live in one dimension higher to the planar matrix fields. The Hamiltonian in that case reads: 
$$Z[J^{m}]=\int \mathcal{D}\alpha(x,z)\mathcal{D}j^{m}(x,z)\mathcal{D}p_{m}(x,z)\times$$
\begin{equation}\times e^{iN^{2}\int\textrm{d}^{D}x \textrm{d}z\sqrt{g} \left(p_{m}(x,z)\frac{\textrm{d}j^{m}(x,z)}{\textrm{d}z}-\alpha(x,z)H_{QRG}(j^{m}(x,z),p_{m}(x,z))\right)}|_{(j^{m}(x,z=0),p_{m}(x,z=0))=(J^{m}(x),\langle O_{m}(x)\rangle)}Z[j^{m}].\end{equation}
The factor of $N^{2}$ out in front of the integral in the exponent plays the role of $\hbar^{-1}$ for the partition function on the space of sources. 

The large $N$ limit is the same as taking the semiclassical limit of the bulk theory which in other words, allows the functional integral to be performed in the saddle point approximation. This saddle point corresponds to extremising the action $S_{B}=\int \textrm{d}^{D}x \textrm{d}z \,(p_{m}\dot{j}^{m}-\alpha(x)H_{QRG}).$ 
The Hamiltonian density takes the form:
$$ H_{QRG}(j^{m}(x,z),p_{m}(x,z))=$$
\begin{equation}\beta^{m}(j^{m}(x,z))p_{m}(x,z)+G^{mn\left\{\mu\right\}\left\{\nu\right\}}\big(j^{m}(x,z)\big)\nabla_{\left\{\mu\right\}}p_{m}(x,z)\nabla_{\left\{\nu\right\}}p_{n}(x,z)+\mathcal{L}_{C}(j^{m}(x,z)).\label{rgham}\end{equation}
The truncation to the single trace subspace is what leads to the quadratic term in the momenta and thus the truncation itself is responsible for the non trivial dynamics of these fields. The potential for the bulk fields is given by the generalisation of the term which renormalizes the coupling of the identity operator: $\mathcal{L}_{C}(j^{m}(x,z))$. 

To conclude, the classical phase space in which QRG flow takes place is thus parameterised by the conjugate pairs $(j^{m}(x,z),p_{m}(x,z))$. This means that they satisfy the fundamental Poisson bracket relation
\begin{equation} \left\{j^{m}(x,z),p_{n}(y,z)\right\}=\delta^{m}_{n}\delta(x,y).\end{equation}

The Hamiltonian generating this QRG flow is $H_{QRG}(j^{m}(x,z),p_{m}(x,z))$ given by \eqref{rgham}. The phase space is thus a sub space (that of single race operators) of the one identified in the previous section, but the Hamiltonian now contains a term quadratic in the momentum. 
\subsection{Revealing the anomalous Ward Identity corresponding to violation of Weyl Invariance}
Before proceeding, the role of the QRG Hamiltonian and the manner in which it encodes coarse graining requires further clarification. in the previous subsection, I mentioned that the Weyl factor is integrated over as a means to encode the summing over all possible RG paths. This also serves to highlight the `path independence' of QRG flow, which corresponds to the freedom to choose a different rate of coarse graining in a space dependent manner at every RG step. In the QRG action, the Weyl factor appears as a Lagrange multiplier, and the functional integral can formally be performed to find:
$$Z[J^{m}]=\int \mathcal{D}j^{m}(x,z)\mathcal{D}p_{m}(x,z)\delta(H_{QRG})\times$$ \begin{equation}e^{iN^{2}\int\textrm{d}^{D}x \textrm{d}z\sqrt{g} p_{m}(x,z)\dot{j^{m}}(x,z)}|_{(j^{m}(x,z=0),p_{m}(x,z=0))=(J^{m}(x),\langle O_{m}(x)\rangle)}Z[j^{m}].\end{equation}
This form of the QRG partition function is particularly illuminating, because it highlights the role of the Hamiltonian in encoding the anomalous Ward identity corresponding to broken Weyl invariance. The delta function in the functional integral imposes this anomalous Ward identity at each and every RG step. 

Another way to see this is to note that at the fixed point, the beta functions vanish and the only term in the Hamiltonian is the potential term, which is the conformal anomaly appearing as a consequence of putting the conformal field theory at the fixed point on an arbitrary background. Away from criticality, the beta functions also contribute to the failure of the theory to maintain Weyl invariance and hence appear alongside the anomaly in the QRG Hamiltonian. This is but a manifestation of the fact that RG flows are triggered by the breaking of conformal symmetry and the beta functions and the anomaly simply measure the response of the effective action to the breaking of this invariance. The anomaly too can be viewed as the beta function encoding the running of the coupling to the identity operator \cite{Dolan}.

In principle, were there other Ward identities in the theory, they too must be included in the QRG path integral formula in an analogous manner\footnote{The holographic dictionary matches Ward identities in the quantum field theory to constraints in the dual gravitational theory, which was noticed in \cite{Corley}}. They should correspond to constraints in the total QRG Hamiltonian and the Lagrange multipliers corresponding to these constraints should be integrated over like the Weyl factor was. The Hamiltonian is thus a total constraint, and this is very reminiscent of the situation in general relativity. 

\subsection{Emergent Gravity from the Quantum Renormalization Group}
The QRG procedure in the context of the matrix field theory will also promote the source of the single trace energy momentum tensor, {\em i.e.} the metric to a dynamical field. As mentioned before, in order to study the pure gravity limit in the bulk,  a limit where the energy momentum tensor is the only operator in the theory with finite scaling dimension needs to be considered. This can happen if all operators acquire large anomalous dimensions but the energy momentum tensor is protected by its Ward Identity. This implicitly also requires the strong coupling on the planar matrix field theory's side, although the classicality of the bulk still requires the large $N$ limit. It must also be assumed that there are no other conserved higher spin currents in the theory. 

The phase space variables in this case will be the metric and the vacuum expectation value of the energy momentum tensor $(\pi_{\mu\nu}(x,z),g^{\mu\nu}(x,z))$, satisfying fundamental Poisson bracket relation
\begin{equation} \left\{\pi_{\mu\nu}(x,z),g^{\alpha\beta}(y,z)\right\}=\delta^{(\alpha}_{\mu}\delta^{\beta)}_{\nu}\delta(x,y).\end{equation}
The bulk RG Hamiltonian then takes the form
\[
\int\textrm{d}^{3}x\sqrt{g}\left(\alpha(x,z)H(\pi_{\mu\nu},g^{\mu\nu})+\xi^{\mu}(x,z)H_{\mu}(\pi_{\mu\nu},g^{\mu\nu})\right)=
\]

\[
\int\textrm{d}^{3}x\sqrt{g}\alpha(x,z)\left(V(g)+\frac{G^{\mu\nu\alpha\beta\left\{ \eta\right\} \{\rho\}}(g)}{2}\nabla_{\left\{ \eta\right\} }\pi_{\mu\nu}\nabla_{\left\{ \rho\right\} }\pi_{\alpha\beta}+\beta^{\mu\nu}(g)\pi_{\mu\nu}\right)+
\]

\begin{equation}
+\int\textrm{d}^{3}x\sqrt{g}\xi^{\mu}(x,z)\left(\nabla^{\nu}\pi_{\mu\nu}\right).
\end{equation}

The first term is the RG Hamiltonian described in the previous section, and it generates local RG transformations. I will denote the Hamiltonian itself, (as opposed to the density) as
\begin{equation} H(\alpha)=\int\textrm{d}^{3}x\sqrt{g}\alpha(x,z)H(\pi_{\mu\nu},g^{\mu\nu}),\end{equation}
and for reasons mentioned in the previous paragraph, this is a constraint with the function $\alpha$ is the corresponding Lagrange multiplier. 
Diffeomorphism invariance
of the matrix field theory arising due to being coupled to an arbitrary background also needs to be taken into account. This is captured by the Ward Identity:

\begin{equation}
\left\langle \nabla^{\mu}T_{\mu\nu}\right\rangle =0,
\end{equation}
that is imposed in the QRG as a constraint:
\begin{equation}
H_{\mu}(\xi^{\mu})=\int\textrm{d}^{3}x\sqrt{g}\xi^{\mu}\left(\nabla^{\nu}\pi_{\mu\nu}\right)=0,
\end{equation}
where the the shift vector $\xi^{\mu}$ is the Lagrange multiplier
enforcing this constraint. 

Thus the phase space of the bulk theory is that of general relativity in Hamiltonian form (discovered by Arnowitt--Deser and Misner in \cite{ADM}). Of course, the algebraic form of the scalar constraint is not quite that of general relativity. The question I will address in the remainder of the article is under what circumstances the QRG scalar constraint becomes that of general relativity. 

The functions $V(g)$ , $\beta^{\mu\nu}(g)$ and $G^{\mu\nu\alpha\beta\left\{ \eta\right\} \{\rho\}}(g)$
are not just functions of the metric but also its derivatives ({\em i.e.}
curvature tensors) to arbitrarily high orders, but they admit a derivative
expansion where the leading order terms are:

\begin{eqnarray}
V(g)=-c_{0}+c_{1}R+\cdots, & \beta^{\mu\nu}(g)=\beta g^{\mu\nu}+\cdots, & G^{\mu\nu\alpha\beta\left\{ \eta\right\} \{\rho\}}(g)=\frac{\gamma}{g}G_{\lambda}^{\mu\nu\alpha\beta}+\cdots.
\end{eqnarray}
 Here, $G_{\lambda}^{\mu\nu\alpha\beta}=g^{\mu\alpha}g^{\nu\beta}+g^{\mu\beta}g^{\nu\alpha}-\lambda g^{\mu\nu}g^{\alpha\beta}$.

If one entertains the possibility of tuning the constants in the gradient expansion then it is conceivable that the ADM scalar constraint can be obtained through the following steps. 
The term linear in the momentum needs to be removed somehow in order to even match the terms in the ADM scalar constraint. Assuming this is done, if the action were re-written by truncating to the first few orders
in the derivative expansions shown, or in other words the constants multiplying the higher order terms are all set to zero. Further, if the relevant constants are tuned take the values $\lambda=1=\gamma=c_{1},\, c_{0}=\Lambda_{cc}$,\footnote{Here, $\Lambda_{cc}$ denotes the Cosmological constant, and shouldn't be confused with the UV cut-off $\Lambda$.}
then the bulk Hamiltonian will be a sum of the Hamiltonian and diffeomorphism
constraints of general relativity. {\em i.e.}
\begin{equation}
H(\pi_{\mu\nu},g^{\mu\nu})=\underset{H_{ADM}(\pi_{\mu\nu},g^{\mu\nu})}{\underbrace{\sqrt{g}(-\Lambda_{cc}+R)+\frac{1}{\sqrt{g}}G_{\lambda=1}^{\mu\nu\alpha\beta}\pi_{\mu\nu}\pi_{\alpha\beta}}}+\cdots.
\end{equation}
 The diffeomorphism constraint is the same as that of general relativity. Note that including any of the terms with spatial derivatives of higher order than those included above in the Hamiltonian will lead to a breakdown of general covariance of the theory. The reason for this breakdown lies in the mismatch between the number of radial gradients hiding in the two powers of the momenta in the kinetic term, and the spatial gradients. One simple way to see why the ADM Hamiltonian has just the right number of these derivatives is to perform the Legendre transform to find that the Lagrangian thus obtained can be written as a scalar density of weight one formed from the spacetime metric (\cite{ADM}).
 
 There isn't any reason {\em a priori} to believe that the aforementioned truncations follow from any of the limits already imposed on the matrix field theory. What would be more satisfying would be to find some additional criterion that the coarse graining mechanism need satisfy from which the ADM form for the constraints follow. 

\section{Consistency conditions of the renormalization group flow and Hamiltonian evolution}
The geometerization of the renormalization
group lies in identifying the radial evolution of a constant `RG time' hypersurface into the bulk with the local quantum renormalization group flow of
the boundary theory. This is only strictly true however, if the evolution is generated by
normal deformations of this hypersurface which satisfy a certain
commutator or Poisson bracket algebra. This algebra can be seen as a consistency condition for the Hamiltonian evolution, because satisfying this condition is necessary for the Hamiltonian flow to not stray away from the sub space of phase space where the constraints are satisfied. And through QRG, it will also reflect the consistency of the local renormalization group flow, in that it dictates how LRG transformations are composed consistently.

The specific form of the structure functions of
this algebra is dictated by the diffeomorphism invariance of the $D+1$
dimensional target space into which this hypersurface is embedded. In other words, the property of the generators of the deformations of hypersurfaces forming a certain commutator or Poisson bracket algebra is a signature that these hypersurfaces are embedded into a one higher dimensional (here Euclidean) spacetime. This algebra is known as the hypersurface deformation algebra or the Dirac algebra. This is because it mirrors the Lie bracket algebra of components of spacetime vector fields decomposed tangentially and orthogonally to an embedded hypersurface \cite{Teitelboim}. 

In this section, the consequences of imposing the Dirac algebra through the Wess--Zumino consistency condition for the Quantum renormalization group will be investigated. From the field theory perspective, this will amount to relating the anomalous Ward identity for broken Weyl/scale invariance to the Ward identity corresponding to diffeomorphism invariance of the boundary theory coupled its background metric at every scale. This relation between said Ward Identities will then impose restrictions on the algebraic form of the RG Hamiltonian, which is nothing but the Hamiltonian of the dual gravity theory and constrain it to be that of general relativity. 

The guarantee that the only representation of this algebra on the gravitational phase space being the ADM constraints follows from a theorem of Hojman, Kuchar and Teitelboim \cite{HKT}. Furthermore, the gradient formula for the metric beta function can also be shown to follow from this demand that the constraint algebra close in a specific form. The key result of interest form which this fact shall follow was first proven by Kuchar in \cite{Kuchar}. 

\subsection{The Wess--Zumino consistency condition as the holographic dual to the Hypersurface deformation algebra}
The Wess--Zumino consistency condition for the local renormalization group is simply a statement of the fact that Weyl transformations commute. This means that when two local RG transformations are composed, it doesn't really matter which of these transformations are performed first, and which is performed second. Consider the the generating functional and focus on its dependence on the metric: $W[g_{\mu\nu}]\equiv\textrm{ln}Z[g_{\mu\nu}]$. The statement of the commutativity of the local RG transformations reads
\begin{equation}[\Delta_{\alpha(x)\delta z},\Delta_{\alpha'(x)\delta z}]W[g_{\mu\nu}(x)]=0.\end{equation}

Even for a conformal field theory on a curved background, this condition imposes non trivial constraints on the form of the conformal anomaly:
\begin{equation}[\Delta_{\sigma(x)},\Delta_{\sigma'(x)}]W_{CFT}[g_{\mu\nu}]=\Delta_{\sigma}\mathcal{A}_{\sigma'}-\Delta_{\sigma'}\mathcal{A}_{\sigma}=0.\end{equation}
Away from the fixed point, when conformal invariance is broken, these consistency conditions necessarily also lead to non trivial relations the beta functions needs to satisfy, in addition to the anomaly terms. 

In the QRG, the beta functions are coded into terms in the Hamiltonian, so the non trivial relations the consistency conditions impose on the beta functions will thus translate into restrictions on the form of the Hamiltonian. In order to see this more concretely, the evaluation of the left hand side of the Wess--Zumino conditions in the QRG context reads:
\begin{equation}[\Delta_{\sigma},\Delta_{\sigma'}]W[g_{\mu\nu}]=\langle[H(\sigma),H(\sigma')]\rangle \underset{N\rightarrow\infty}{\rightarrow}\langle\left\{H(\sigma),H(\sigma')\right\}\rangle, \end{equation}
 the vanishing of this, is how the Wess--Zumino consistency conditions are encoded in the QRG. 
 The diffeomorphism Ward identity does however allow for the possibility that the right hand side of the action of the commutator of the generators of Weyl transformations to vanish as a consequence of being proportional to the covariant divergence of the energy momentum tensor. From the QRG perspective, this means that the right hand side of the bracket between $H(\sigma)$ and itself (smeared with a different lapse multiplier) can in principle be proportional to the constraint $H_{\mu}$ with some smearing perhaps containing the derivatives of the lapse multipliers. This means that the Poisson algebra of the constraints, particularly a specific form of said algebra is the holographic dual to the Wess--Zumino consistency conditions. 

I conjecture that the anomalous Ward identity corresponding to the broken Weyl or scale invariance of the theory which the Wess--Zumino consistency conditions pertain is, in a specific way, related to the Ward identity corresponding to the diffeomorphism invariance of the theory.\footnote{The mathematical statement of which is the covariant conservation of the energy momentum tensor's vacuum expectation value.} This means that the relationship between these Ward identities imply a specific form of the Poisson algebra of the corresponding dual constraints. This form of the Poisson algebra, given other assumptions I will further mention, will be sufficiently strong to fix the algebraic form of the scalar Hamiltonian constraint to be identical to that of general relativity. 
\subsection*{Kinematics of Hypersurface Deformations}
To start, it will help to describe the hypersurface deformation algebra at the kinematical level. Consider an infinitesimal spacetime diffeomorphism generated by the vector field $v^{a}$\footnote{I will use lowercase latin letters such as $a,b,...$ for $D+1$ dimensional spacetime tensors, which will run form $0$ to $D$.}, {\em i.e.} $y^{a}\rightarrow y^{a}+v^{a}$, it can be decomposed into components tangential and orthogonal to any given hypersurface as
\begin{equation}v^{a}=\sigma n^{a}+v_{\parallel}^{a},\end{equation}
where $\sigma=n^{a}v_{a}$, and $v_{\parallel}^{a}=-(n^{c}v_{c})n^{a}+v^{a}$. Note that this vector is purely tangential to the hypersurface, because $n_{a}v^{a}_{\parallel}=0$, so I will denote it as $v^{\mu}_{\parallel}$ in what follows. The vector $n^{a}$ is the normal to a co-dimension one hypersurface $\Sigma$. The deformation of the hypersurface generated by the vector field $v^{a}$ is given through the action of the operator
\begin{equation}X(v)=\int_{\Sigma} \textrm{d}^{D}x\sqrt{g} v^{a}\frac{\delta}{\delta y^{a}},\label{sdiff}\end{equation}
which satisfies the commutation relations
\begin{equation}[X(v),X(w)]=X([v,w]).\end{equation}
Here $[v,w]$ is the Lie bracket of the vector fields $v^{a}, w^{a}$. Then a foliation dependent decomposition of the above operator can be introduced as follows:
\begin{equation}N_{\sigma}=\int_{\Sigma}\textrm{d}^{D}x\sqrt{g} \sigma n^{a}\frac{\delta}{\delta y^{a}},\end{equation}
\begin{equation}T_{v_{\parallel}}=\int_{\Sigma}\textrm{d}^{D}x \sqrt{g}v^{\mu}_{\parallel}\partial_{\mu}y^{a}\frac{\delta}{\delta y^{a}}. \end{equation}
The algebra of these deformations is given by
\begin{equation}[N_{\sigma},N_{\sigma'}]=-T_{f(\sigma,\sigma')},\, \,\,[T_{v_{\parallel}},N_{\sigma}]=-N_{v_{\parallel}^{\mu}\partial_{\mu}\sigma},\,\,\, [T_{v_{\parallel}},T_{w_{\parallel}}]=T_{[v_{\parallel},w_{\parallel}]}. \label{hdefalg}\end{equation}
Here $f^{\nu}(\sigma,\sigma')=g^{\mu\nu}(\sigma'\partial_{\mu}\sigma-\sigma\partial_{\mu}\sigma').$

It is interesting to see that the above Lie bracket algebra is not a Lie algebra because the analogue of the structure constants, are now replaced by phase space dependent functions, {\em i.e.} the vector $f^{\nu}(\sigma, \sigma')$. It is still analogous to a Lie algebra in the sense that the commutator of these deformations generators closes to other deformation generators. Also, the structure functions of the above deformation algebra are fixed by the demand that when the normal and tangential deformations are combined to form the overall deformation $X(v)$, it satisfied the algebra of spacetime diffeomorphisms \eqref{sdiff}. In other words, the specific form of the structure functions of this algebra are fixed by the demand for full diffeomorphism invariance of the spacetime into which the hypersurface is embedded. This algebra must be mirrored by the Poisson algebra of the constraints on the phase space of any dynamical theory which respects full diffeomorphism invariance. This fact that spacetime structure is reflected in the algebra of constraints was described in \cite{Teitelboim}. 

\subsection*{Implications for gravity}
The result of key importance in the context of the phase space of general relativity is that of Hojman, Kuchar and Teitelboim (HKT) in \cite{HKT}. They prove that the unique representation of the algebra of hypersurface deformations \eqref{hdefalg} on the phase space spanned by the metric on a hypersurface and its conjugate momentum is given by the following constraints:
$$N_{\sigma}\rightarrow \int_{\Sigma}\textrm{d}^{D}x \sigma(x)\left(\frac{1}{\sqrt{g}}\left(\pi_{\mu \nu}\pi^{\mu\nu}-\frac{1}{D-1}\textrm{tr}\pi^{2}\right)-\sqrt{g}(\Lambda_{cc}-R)\right),\,\,\, $$\begin{equation}T_{v_{\parallel}}\rightarrow \int_{\Sigma}\textrm{d}^{D}x \sqrt{g}v^{\nu}_{\parallel}(x)(\nabla_{\mu}\pi^{\mu}_{\nu}). \end{equation}

The Poisson algebra of these constraints mirrors the algebra of hypersurface deformations. These constraint functions are easily recognised as the ADM scalar and diffeomorphism constraints where the lapse Lagrange multiplier is identified with $\sigma(x)$ and the shift multiplier is identified with $v^{\mu}_{\parallel}(x)$. 
Thus the following Poisson algebra being satisfied by the constraints:
$$\left\{H(\sigma),H(\sigma')\right\}=-H_{\mu}(f^{\mu}(\sigma,\sigma')),\,\left\{H_{\mu}(\xi^{\mu}),H(\sigma)\right\}=H(\xi^{\mu}\partial_{\mu}\sigma), \, $$ \begin{equation}\left\{H_{\mu}(\xi^{\mu}),H_{\nu}(\zeta^{\nu})\right\}=H_{\mu}([\xi,\zeta]^{\mu}).\label{hypalg}\end{equation}
is a necessary and sufficient condition for these constraints to take the ADM form. 

The third of the above Poisson bracket relations is a representation of the algebra of spatial diffeomorphism algebra. The second bracket is entirely a consequence of the fact that the scalar constraint density is a tensor density of weight one. In a sense, these brackets pertain to just kinematics, as far as QRG is concerned. This is because the diffeomorphism constraint in the total QRG Hamiltonian is already of the same algebraic form as the diffeomorphism constraint of GR, and hence necessarily satisfies the same Poisson algebra. Also, the tentative scalar constraint density being a tensor density of weight one only instructs where factors of $\sqrt{g}$ ought to appear in each of its terms, but it doesn't fix the functional dependence of the functions themselves on the phase space variables. The first Poisson bracket relation however does indeed pertain to dynamics. For it to be satisfied, the form of various functions in the scalar constraint are fixed.  

Going back to the quantum renormalization group, there is now potential to impose a condition on the very coarse graining scheme itself, the satisfaction of which will force the RG Hamiltonian to take the ADM form. This condition is the holographic dual to the hypersurface deformation algebra. It is given by:
\begin{equation}[\Delta_{\sigma},\Delta_{\sigma'}]W[g_{\mu\nu}]=\int \textrm{d}^{3}x \sqrt{g} f^{\mu}(\sigma,\sigma')\langle\nabla_{\rho}T^{\rho}_{\mu}\rangle,\end{equation}
which, as mentioned before is but a particular manner in which the Wess Zumino consistency condition is satisfied (because $\langle\nabla_{\rho}T^{\rho}_{\mu}\rangle=0$). 

I will now describe how the functions in the QRG Hamiltonian can be fixed by demanding this particular form of the consistency condition, starting with the kinetic term. 

\subsection{The Kinetic term}
The tentative kinetic term is one which is quadratic in the momentum, but only the first term in the gradient expansion is ultra local in the metric and momenta. The QRG scalar constraint in this case takes the form
\begin{equation}H(\sigma)=\int_{\Sigma}\textrm{d}^{D}x \alpha(x,z)\left(\frac{1}{\sqrt{g}}\left(\pi_{\mu \nu}\pi^{\mu\nu}-\frac{1}{D-1}\textrm{tr}\pi^{2}\right)+F(g,\pi)-\beta_{\mu\nu}(g)\pi^{\mu\nu}+\sqrt{g}V(g)\right).\end{equation}

The function $F(g,\pi)$ stands for the rest of the terms in the gradient expansion of the quadratic in momentum term. This is a function which consists of an arbitrary number of derivatives of the metric and the two powers of momenta. 
The key Poisson bracket relation to use in order to fix the form of the remaining functions in this constraint is the bracket between two scalar constraints, {\em i.e.}
\begin{equation}\left\{H(\alpha),H(\alpha')\right\}=H_{\mu}(f^{\mu}(\alpha,\alpha')),\label{da}\end{equation}
where leaving aside the specific form of $f^{\mu}(\alpha,\alpha')$, a lot can be gained from just noticing the fact that the right hand side of the above bracket is linear in the momentum. In order to exploit this feature, it is useful first to recall that the Poisson brackets between two phase space functions reduce the total polynomial order in the momenta by one and maintain the order of spatial derivatives. The bracket between two scalar constraints breaks up into a sum of several terms which can be ordered based on the total polynomial order of the momenta. The highest order term from this counting would be the bracket between the kinetic term and itself:
$$\left\{\int_{\Sigma}\textrm{d}^{D}x\alpha(x,z)\left(\frac{1}{\sqrt{g}}G^{\mu\nu\alpha\beta}\pi_{\mu\nu}\pi_{\alpha\beta}+F(g,\pi)\right),\int_{\Sigma}\textrm{d}^{D}y\alpha'(y,z)\left(\frac{1}{\sqrt{g}}G^{\mu\nu\alpha\beta}\pi_{\mu\nu}\pi_{\alpha\beta}+F(g,\pi)\right)\right\}$$ \begin{equation}-(\alpha\leftrightarrow \alpha'),\end{equation}
This will split up again into three terms, the first being
\begin{equation}\left\{\int_{\Sigma}\textrm{d}^{D}x\frac{\alpha(x,z)}{\sqrt{g}}G^{\mu\nu\alpha\beta}\pi_{\mu\nu}\pi_{\alpha\beta},\int_{\Sigma}\textrm{d}^{D}y\frac{\alpha'(y,z)}{\sqrt{g}}G^{\mu\nu\alpha\beta}\pi_{\mu\nu}\pi_{\alpha\beta}\right\}-(\alpha\leftrightarrow \alpha')\end{equation}
which vanishes because of the ultra locality of the resulting expression and subsequent anti-symmetrisation of the smearing functions. 
The second and third terms of the form
\begin{equation}\left\{\int_{\Sigma}\textrm{d}^{D}x\frac{\alpha(x,z)}{\sqrt{g}}G^{\mu\nu\alpha\beta}\pi_{\mu\nu}\pi_{\alpha\beta},\int_{\Sigma}\textrm{d}^{D}y \alpha'(y,z) F(g,\pi)\right\}-(\alpha\leftrightarrow \alpha'),\end{equation}
\begin{equation}\left\{\int_{\Sigma}\textrm{d}^{D}x \alpha(x,z)F(g,\pi),\int_{\Sigma}\textrm{d}^{D}y \alpha'(y,z) F(g,\pi)\right\}-(\alpha\leftrightarrow \alpha'),  \label{2manyder}\end{equation}
don't however identically vanish due to the presence of spatial gradients, and thus leads to a set of terms genuinely cubic in the momenta. 
These terms vanish if and only if $F(g,\pi)=0$. The study of such terms is the subject of \cite{me}. In that work, the demand that the structure functions be independent of the momenta even is not imposed, but still, the constraint algebra doesn't remain first class under a wide class of such modifications of the kinetic term. This eliminates the potentially cubic term in the result of this Poisson bracket relation.

Thus from just positing this holographic version of the Wess--Zumino consistency condition, it follows that 
\begin{equation}G^{\mu\nu\alpha\beta\left\{\eta\right\}\left\{\rho\right\}}\rightarrow \frac{1}{\sqrt{g}}G^{\mu\nu\alpha\beta}=\frac{1}{\sqrt{g}}(g^{\mu(\alpha}g^{\beta)\nu}-g^{\mu\nu}g^{\alpha\beta}).\end{equation}
Thus the double trace beta function is an ultra local function of the metric known as the de-Witt super metric. This also implies that the kinetic term of the Hamiltonian is ultra local in both the metric and the momenta. This is a canonically normalised kinetic term, akin to that which is encountered in most field theories. 

The super metric, being a metric on the space of metrics is used to define an inner product on field space, which in turn is necessary to define the functional integral over geometries in the quantum theory. Paraphrasing from a discussion in \cite{Mottola}, allowing this inner product to be taken with respect to a super metric containing derivatives of the metric would have the effect of defining a different set of dynamical fields in the theory. So the specification of an ultra local super metric can also be seen as a manifestation of the fact that the metric is taken to be the fundamental field variable and that the presence of derivatives of it in the action is the cause for dynamics. 

Furthermore, the ultra locality of the kinetic term in both the metric and the momenta will ensure the invertibility of the relationship between the canonical momenta and the extrinsic curvature tensor. The extrinsic curvature tensor is defined as 
\begin{equation}K_{\mu\nu}=-\frac{1}{2}\mathcal{L}_{n}g_{\mu\nu},\end{equation}
where $n^{\mu}$ is the vector normal to the hypersurface. The relationship between this tensor and the canonical momentum given the ultra local kinetic term is 
\begin{equation}K_{\mu\nu}=\frac{1}{\sqrt{g}}\left(\pi_{\mu\nu}-\frac{1}{D-1}\textrm{tr}\pi g_{\mu\nu}\right).\end{equation}
The simple algebraic nature of the relationship between the canonical momenta and the extrinsic curvature is one of the many necessary conditions to find a Lagrangian that can be rewritten in the form of the Einstein Hilbert action which is manifestly covariant in $D+1$ dimensions after performing the Legendre transform. 

The HKT result also seems to demand that the metric beta function ought to simply vanish in order for the algebra of constraints to be satisfied, and also, the potential term should be truncated to just the first two terms in its derivative expansion. There is a subtlety here regarding the fate of the term linear in the momentum and the potential term and it will be elucidated and addressed in the subsection to follow.  
\subsection{Gradient flow formula for the metric beta function and canonical transformations}

The constraint algebra enforced through the holographic Wess--Zumino consistency conditions also has implications for the form of the beta function and potential term in the RG Hamiltonian. 

In the last section, it was deduced that the kinetic term should be ultra local in both the metric and the momenta in order to satisfy the hypersurface deformation algebra. Assuming only this, the scalar constraint is given by
\begin{equation}H(\alpha)=\int_{\Sigma}\textrm{d}^{D}x \alpha(x,z)\left(\frac{1}{\sqrt{g}}\left(\pi_{\mu \nu}\pi^{\mu\nu}-\frac{1}{D-1}\textrm{tr}\pi^{2}\right)-\beta_{\mu\nu}(g)\pi^{\mu\nu}+\sqrt{g}V(g)\right).\end{equation}
I will now sketch the derivation of the result showing that the demand that the hypersurface deformation algebra be satisfied will translate into the so called `gradient formula' for the metric beta function. For more detailed computations proceeding along this line of reasoning to derive this result, see \cite{me}. The original derivation of this result came from an effort to formulate the HKT theorem in the Lagrangian framework by Kuchar in \cite{Kuchar}.

Given that the vanishing of the cubic resulting from the Poisson brackets has been established, the next to higher order term will be a quadratic in momentum expression which comes from the bracket between the kinetic term and the term linear in the momentum:
\begin{equation}\left\{\int_{\Sigma}\textrm{d}^{D}x\frac{\alpha(x,z)}{\sqrt{g}}G^{\mu\nu\alpha\beta}\pi_{\mu\nu}\pi_{\alpha\beta},\int_{\Sigma}\textrm{d}^{D}y\alpha'(y,z) \beta_{\mu\nu}(g)\pi^{\mu\nu}\right\}-(\alpha\leftrightarrow \alpha').\end{equation}
The function $\beta^{\mu\nu}(g)$ depends on the metric and its momenta, and the above expression will not identically vanish despite the anti-symmetrisation of the smearing functions. In order to satisfy the hypersurface deformation algebra however, this expression must strongly vanish, by virtue of the fact that there is no term on the right hand side of \eqref{da} that is quadratic in the momenta. 

 It can be shown \footnote{See section 5 of \cite{me}} that the vanishing of the above quadratic in momentum term will imply that
\begin{equation}\frac{\delta(G^{\alpha\gamma\mu\nu}\beta_{\alpha\gamma})(x)}{\delta g_{\rho\eta}(y)}-\frac{\delta(G^{\delta\kappa\rho\eta}\beta_{\delta\kappa})(y)}{\delta g_{\mu\nu}(x)}=0.\label{funccurl}\end{equation}
Following similar logic, a term generated by the Poisson bracket calculation that is momentum independent should also vanish. The relevant piece of the bracket here will be
\begin{equation}\left\{\int_{\Sigma}\textrm{d}^{D}x\alpha(x,z) \beta_{\mu\nu}(g)\pi^{\mu\nu},\int_{\Sigma}\textrm{d}^{D}y\sqrt{g}\alpha'(y,z) V(g)\right\}-(\alpha\leftrightarrow \alpha'),\end{equation}
which produces a term that is independent of the momenta and can be split up into a sum of terms ordered by the number of spatial gradients. The first non-trivial order of derivatives will be the second order, the vanishing of which is 
\begin{equation}\nabla_{\gamma}(G^{\alpha\gamma\mu\nu}\beta_{\mu\nu})=0.\label{divfree} \end{equation}
The conditions \eqref{funccurl} and \eqref{divfree} then imply that the function $\beta_{\mu\nu}(g)$\footnote{for a sketch of the proof of this statement, see \cite{Kuchar} and references therein} has to take the form
\begin{equation}\beta_{\mu\nu}(g)=G_{\mu\nu\rho\eta}\frac{\delta c[g]}{\delta g_{\rho\eta}},\label{gradform}\end{equation}
where $c[g]$ is a functional of the metric and its derivatives. 

From the quantum field theory perspective, the equation \eqref{gradform}, is the so called gradient formula for the metric beta function. This result was arrived at solely through considerations of the (`holographic') Wess--Zumino consistency conditions much akin to how such a formula is derived in the traditional local RG literature, for instance in \cite{Osborn1}, \cite{Freidan1}. Such a formula was also derived from considerations of entanglement entropy in holographic theories in \cite{GRG}, in the case where $c[g]$ takes the form of the Einstein Hilbert action.

Coming back to the dual gravitational theory, consider just the kinetic term and the term linear in the momentum, the sum of which can be manipulated as follows
$$
\frac{1}{\sqrt{g}}G_{\mu\nu\rho\eta}\pi^{\mu\nu}\pi^{\rho\eta}-G_{\mu\nu\rho\eta}\frac{\delta c[g]}{\delta g_{\rho\eta}}\pi^{\mu\nu}
$$
\begin{equation}=\frac{1}{\sqrt{g}}G_{\mu\nu\rho\eta}\left(\pi^{\mu\nu}-\frac{1}{2}\frac{\delta c[g]}{\delta g_{\mu\nu}}\right)\left(\pi^{\rho\eta}-\frac{1}{2}\frac{\delta c[g]}{\delta g_{\rho\eta}}\right)-\frac{1}{4}G_{\mu\nu\rho\eta}\frac{\delta c[g]}{\delta g_{\mu\nu}}\frac{\delta c[g]}{\delta g_{\rho\eta}}.
\end{equation}

This manipulation makes the possibility for the following canonical transformation
$$\pi^{\mu\nu}\rightarrow \pi^{\mu\nu}-\frac{1}{2}\frac{\delta c[g]}{\delta g_{\mu\nu}},$$
apparent. This is a canonical transformation because it preserves the canonical Poisson brackets of the theory and subsequently comes at the cost of adding a total derivative term to the action. The role of such canonical transformations in holographic RG was discussed in detail in \cite{Papa1}. It follows from this canonical transformation that 
$$\int \textrm{d}z\int_{\Sigma}\textrm{d}^{D}x \sqrt{g}\pi^{\mu\nu}\dot{g}_{\mu\nu}\rightarrow \int \textrm{d}z \int_{\Sigma}\textrm{d}^{D}x \sqrt{g}\pi^{\mu\nu}\dot{g}_{\mu\nu}+\int\textrm{d}z\int_{\Sigma}\textrm{d}^{D}x \sqrt{g}\frac{1}{2}\frac{\delta c[g]}{\delta g_{\mu\nu}}\dot{g}_{\mu\nu}$$
\begin{equation}=>\int \textrm{d}z\int_{\Sigma}\textrm{d}^{D}x\sqrt{g} \pi^{\mu\nu}\dot{g}_{\mu\nu}+c[g]|^{z=z_{*}}_{z=0}.\end{equation}
This is just the statement of the fact that $c[g]$ is the generating functional of the aforementioned canonical transformation. 

This effectively removes the linear term in the scalar constraint, leaving only the ultra local, canonical kinetic term. The last term in the Hamiltonian constraint whose form hasn't yet been fixed in the above discussion from the demand of satisfaction of the hypersurface deformation algebra is the momentum independent potential term. It too gets modified as a consequence of the canonical transformation mentioned above, {\em i.e.}
\begin{equation}
V(g)\rightarrow V(g)-\frac{1}{4}G_{\mu\nu\rho\eta}\beta^{\mu\nu}(g)\beta^{\rho\eta}(g)\equiv U(g).
\end{equation} 

Now, the form of the Hamiltonian constraint density after the canonical transformation is:
$$\frac{1}{\sqrt{g}}G_{\mu\nu\rho\eta}\pi^{\mu\nu}\pi^{\rho\eta}+U(g),$$
and the HKT result will lead to the condition 
\begin{equation}U(g)=V(g)-\frac{1}{4}G_{\mu\nu\rho\eta}\beta^{\mu\nu}(g)\beta^{\rho\eta}(g)=\sqrt{g}(-\Lambda_{cc}+R).\end{equation}

This difference between the potential term and the square of the beta function being exactly the potential term in the ADM Hamiltonian constraint is also related to the vanishing of the difference between the $a$ and $c$ anomaly coefficients in AdS/CFT as was discussed in \cite{NakayamaLRG}.\footnote{It should be noted that conventions to do with factors of 2 differ between this article and \cite{NakayamaLRG}.}

I wish to emphasise that the above subsection provides a {\em derivation} of the gradient condition for the metric beta function which was so far assumed in discussions relating to the quantum renormalization group. The additional input however was the holographic Wess--Zumino consistency conditions. On the gravity side of the duality, these conditions translate into closure of the constraint algebra in a very specific manner, which is a stronger condition than just the demand for closure of the constraint algebra which was already presented in \cite{S.S.Lee2}.

\subsection{The realm of possibilities}
Despite the many arguments made in the previous sections to justify the conjectured form of the Wess--Zumino consistency conditions, one can nevertheless ask what other consistent choices could have been made on the gravity side for these conditions to be satisfied. A consistent choice of the way in which the Wess--Zumino conditions are satisfied translates through the duality into a manner in which the Poisson algebra of constraints can close. Then, the most general condition one can impose on the bracket between two scalar constraints is just closure {\em i.e.} to require
\begin{equation}\left\{H(\sigma),H(\sigma')\right\}\approx 0.\end{equation}
The symbol $\approx$ denotes ``weak equality'' which means equality when the constraints are satisfied. That would leave the possibility for the Poisson brackets to result in terms proportional to both the scalar and vector constraint with arbitrary structure functions, whose phase space dependence is made explicit with the notation $\tilde{f}^{\mu}(g,\pi;\sigma,\sigma')$, $\tilde{h}(g,\pi;\sigma,\sigma')$:
\begin{equation}\left\{H(\sigma),H(\sigma')\right\}=H(\tilde{h}(g,\pi;\sigma,\sigma'))+H_{\mu}(\tilde{f}^{\mu}(g,\pi;\sigma,\sigma')).\end{equation}
Before proceeding further in this discussion, it will help to first take a step back and recall some basic notions in the theory of constrained Hamiltonian systems. The fact that the Poisson algebra of the constraints results in terms proportional to the constraints themselves implies that the constraint algebra is {\em first class}. The first class nature of the $D+1$ constraints in the general relativity context is the manifestation of the fact that there are $\frac{(D+1)(D-2)}{2}$ true degrees of freedom of the gravitational field. 

Now, going back to the situation of interest, one can ask what class of gravitational theories (i.e. theories defined on the phase space of GR) exist that possess spatial ({\em i.e.} in the field theory's space directions) diffeomorphism invariance and a local quadratic in momentum Hamiltonian constraint which is first class and hence propagate the same number of degrees of freedom as general relativity. No such theory has been found so far, although a complete proof of the statement that no such theory could possibly be found doesn't exist at the moment. Nevertheless, if additional restrictions such as the demand that the kinetic term be ultra local are imposed, then the mere demand for closure of the constraint algebra will force the `tentative' constraints to take the form of those of general relativity, as was shown in \cite{Martinec}. This remains true if the kinetic term is also modified by the addition of an arbitrary local, but quadratic in momentum term, see \cite{me}. If the demand that the modifications no longer remain quadratic in momenta is relaxed, then perhaps there is a wide range of generalisations of the hypersurface deformation algebra that are admissible, such as those described in \cite{Bojowald}. 

If no such theory exists, then what one would conclude is that the {\em only} realisation of a first class scalar constraint in an otherwise spatially covariant theory of gravity is necessarily the ADM Hamiltonian constraint of general relativity. In that case, the only demand that one need impose is for the Wess--Zumino conditions to {\em somehow} be satisfied, {\em i.e.} that the algebra of constraint simply be first class and the only consistent manner in which such closure can be achieved is if the constraints are those of general relativity. Here, there will be no need to make any conjecture about the specific manner in which the Wess--Zumino consistency conditions are satisfied, and the covariance of the dual theory will follow solely from the abelian nature of the group of local Weyl transformations. 
\section{Discussion and Outlook}

I have argued that in order for the quantum renormalization group applied to matrix field theories to yield General relativity in the bulk, the additional imposition of what I call the holographic Wess--Zumino consistency condition is necessary. If a local RG scheme can satisfy a consistency condition of the form posited here, then there are guarantees on the gravity side that the constraints whose Poisson algebra must mirror said condition have to take the ADM form. The most non trivial consequence that follows from this algebra is that the term linear in the momentum generated in the RG Hamiltonian, which in principle breaks RG time reversal invariance, can be removed though a canonical transformation due to the fact that the beta function must take the gradient form. 

That being said, on the field theory side, the problem of finding the local RG scheme which satisfies the holographic Wess--Zumino consistency condition has not yet been addressed, and there isn't any guarantee that such a scheme should exist. It should be noted however that in the context of higher spin holography, the authors in \cite{LeighRG} find a scheme that guarantees bulk diffeomorphism invariance. Addressing this issue will be the subject of future work. 

There is another caveat to be made regarding the holographic Wess-Zumino consistency condition. In the presence of asymptotic boundaries, the gravitational Hamiltonian needs to have boundary terms, added to it, for consistency of the variational principle (see \cite{ReggeTeitelboim1}, \cite{ReggeTeitelboim2}). These boundary terms may also lead to central extension of the constraint algebra when the Poisson bracket algebra of the appended constraints is considered. This is the case for instance when asymptotically dS or AdS spacetimes are considered.
Such a treatment of the asymptotically AdS boundary and is related to Holographic Renormalization through which the computation of the conformal anomaly can be performed, see for instance \cite{Skenderis}, \cite{Imbimbo1}. Asymptotically AdS boundary conditions can be specified through fall off conditions of the gravitational phase space variables (which is intimately related to the Fefferman--Graham expansion), and the generators of the group of asymptotic symmetries are the bulk diffeomorphisms restricted to respect said conditions. Among such diffeomorphisms, there are those whose action reduces to Weyl transformations of the boundary metric, and these are known as as PBH (Penrose, Brown--Hennaux) transformations in the literature. In \cite{Erd}, the case was made for equating local RG transformations to these PBH transformations and this was seen as the field theoretic description of holographic RG flow. Ideally, a suitable local RG scheme should also satisfy the centrally extended version of the holographic Wess-Zumino consistency condition. This might be how QRG  is bridged with other approaches to the holographic renormalisation group that focus on radial evolution in the vicinity of the asymptotic boundary, as described for instance in \cite{HRG}. It should be mentioned however that there are more manifestly covariant coarse graining mechanisms to delve into the bulk rather than just to radially evolve the boundary, as was described in \cite{NetEng}. 

Another interesting question would be to see what happens when no assumption is made about the spectrum of scaling dimensions of the various operators of the matrix field theory, and whether the more general gradient formula for such operators can be derived from demanding consistency of the renormalization group flow. The sources for the operators will be promoted to dynamical higher spin fields, and if the gradient formula is satisfied, then it becomes clear that these fields are massive and their masses are intimately related to their scaling dimensions (\cite{S.S.Lee1}).

 From studies of the open/closed string duality, which AdS/CFT is seen to be but a manifestation of, there is an expectation that the holographic metric beta function for instance should be related to the beta function corresponding to the background metric of the non linear sigma model representation of the string. This was discussed in the conclusion section of \cite{KhouryVerlinde} for instance.  

There is also the possibility that the QRG can somehow catch a glimpse of a non linear sigma model. In this case, such a relationship might follow form the fact that if no particular asymptotics are assumed, then the generating functional for the canonical transformation which removes the linear in momentum term in the Hamiltonian constraint will be the sole contributor to the on shell action or Hamilton--Jacobi functional\footnote{the asymptotic counter-terms can also be subsumed into this generating functional, see \cite{Papa1}, \cite{Papa2}}. In this situation, the gradient flow formula for the beta function on shell is very much like the equation defining the beta functions in a non linear sigma model \cite{Tseytlin}. To make this more than an analogy, one needs to introduce a scalar dilaton field which compensates for the Weyl non-invariance of the action and thereby acts as a Stuckleberg field. Should this be appropriately introduced, then the on shell action can presumably be related to the low energy effective action of the sigma model action and the dilaton--gravity Hamilton--Jacobi equations should coincide with the beta function equations describing the consistency conditions for a non-critical string propagating in the bulk. The emergence of the extra dimension in this situation is a consequence of world sheet conformal symmetry being anomalous. This will warrant the introduction of the Liouville mode and when the right choice of integration measure is made, this will be counted among the other scalar fields or target space co-ordinates. Holographic renormalization can be seen through the shift of the Liouville zero mode integration bounds \cite{DW}. 

The above paragraph is but a sketch of ideas and expectations which warrant further investigation in order to be made concrete. Investigations along these lines will be the subject of future research. 
\section*{Acknowledgments}
I would like to thank Sung-Sik Lee for educating me about the Quantum Renormalization Group through several discussions. I thank Tim Koslowski for asking the question of wether the gradient flow property must be imposed or can be deduced from other features of the QRG flow. I thank Lee Smolin for carefully reading through earlier versions of this manuscript and providing very useful comments and Henrique Gomes and Yasha Neiman for many useful discussions. I would also like to thank Yu Nakayama for enlightening email exchanges. 

This research was supported in part by Perimeter Institute for Theoretical Physics as well as by grant from NSERC and the John Templeton Foundation. Research at
Perimeter Institute is supported by the Government of Canada through the Department of Innovation,
Science and Economic Development Canada and by the Province of Ontario through the Ministry of
Research, Innovation and Science
 
\end{document}